\documentstyle[aps,twocolumn,amssymb]{revtex}
\begin{document}

\draft
\preprint{hep-th/9804001, UCR--CIMM--4--98}

\title{On the ultraviolet behaviour of
       quantum fields over noncommutative manifolds}
\author{Joseph C. V\'arilly}
\address{Department of Mathematics, Universidad de Costa Rica,
         San Pedro 2060, Costa Rica}
\author{Jos\'e M. Gracia-Bond{\'\i}a}
\address{Department of Physics, Universidad de Costa Rica,
         San Pedro 2060, Costa Rica}
\date{\today}
\maketitle
\begin{abstract}
By exploiting the relation between Fredholm modules and the
Segal--Shale--Stinespring version of canonical quantization, and
taking as starting point the first-quantized fields described by
Connes' axioms for noncommutative spin geometries, a Hamiltonian
framework for fermion quantum fields over noncommutative manifolds is
introduced. We analyze the ultraviolet behaviour of second-quantized
fields over noncommutative 3-tori, and discuss what behaviour should
be expected on other noncommutative spin manifolds.
\end{abstract}


\newbox\ncintdbox \newbox\ncinttbox 
\setbox0=\hbox{$-$}
\setbox2=\hbox{$\displaystyle\int$}
\setbox\ncintdbox=\hbox{\rlap{\hbox
    to \wd2{\hskip-.125em\box2\relax\hfil}}\box0\kern.1em}
\setbox0=\hbox{$\vcenter{\hrule width 4pt}$}
\setbox2=\hbox{$\textstyle\int$}
\setbox\ncinttbox=\hbox{\rlap{\hbox
    to \wd2{\hskip-.125em\box2\relax\hfil}}\box0\kern.1em}

\newcommand{\ncint}{\mathop{\mathchoice{\copy\ncintdbox}%
                      {\copy\ncinttbox}{\copy\ncinttbox}%
					  {\copy\ncinttbox}}\nolimits}

\newcommand{\snorm}[1]{\mathopen{|\mkern-2mu|\mkern-2mu|}{#1}%
                       \mathclose{|\mkern-2mu|\mkern-2mu|}}

\section{Introduction}

This article considers quantum fields over noncommutative spaces. The
fact that compactification of matrix models in M-theory leads to
noncommutative tori~\cite{ConnesDS,LizziS} provides some motivation.
But here we address questions of principle, open since Connes
characterized the noncommutative manifolds able to sustain
matter~\cite{ConnesAxioms}.

First quantized fermion fields live on noncommutative spin manifolds,
in particular NC tori. An odd spin geometry consists of four objects
$({\cal A},{\cal H},J,D)$, where: (1)~$\cal A$ is a unital
pre-$C^*$-algebra; (2)~${\cal H}$ is a Hilbert space carrying a
representation of~${\cal A}$ by bounded operators; (3)~$J$ is an
antilinear isometry of ${\cal H}$ onto itself; (4)~$D$ is a
selfadjoint operator on~${\cal H}$, with compact resolvent. From such
a structure, plus some appropriate compatibility conditions formulated
as axioms, Connes was able to derive ordinary spin geometry ---in
which $D$ is the standard Dirac operator $D\mkern-11.5mu/\,$---
including all of the Riemannian structure. Leaving out the condition
that ${\cal A}$ be commutative, we are left with a handle on the vast
new realm of noncommutative spin geometries.

Noncommutative geometry is also a language of choice for the formal
aspects of quantum field theory. For instance, Wick ordering is
intimately related to Connes' Fredholm
modules~\cite{ConnesIHES,ConnesGNC,Book}, reviewed here. The structure
of anomalies in gauge field theories can be recast in terms of cyclic
cohomology; this was pointed out by Araki~\cite{ArakiB,ArakiS} and put
forward by Mickelsson and Langmann in a splendid series of
papers~\cite{MLSaga}. Very recently, it has been found that the
quasi-Hopf algebra structure of Feynman graphs~\cite{Kreimer} is
directly related to Hopf algebras relevant to the general index
formula in noncommutative geometry~\cite{ConnesMHopf}.

These facts can be better put into perspective by taking the step
proposed in this paper. Indeed, it has long been known that quantum
field theory possesses an algebraic core independent of the nature of
space-time~\cite{BaezSZ}. For instance, the description of fermions
coupled to external gauge fields is a problem in representation theory
of the infinite dimensional orthogonal group. From the latter, with
the input of an appropriate single-particle space, it is possible to
derive all quantities of interest: current algebra, anomalous
transformation terms, Feynman rules~\cite{Rhea}. Now, the process is
fundamentally unchanged if the ``matter field'' evolves on a
noncommutative space. In a nutshell: we endeavour to apply the
canonical quantization machinery to a noncommutative kind of
single-particle space.

We couple our proposal here with a description of the simplest
imaginable model, generalizing the textbook field quantization with
periodic boundary conditions; i.e., we quantize chiral fermions in a
3-dimensional ``noncommutative box''. (We readily admit to a lingering
prejudice in favour of the physical number of dimensions.)

The ultraviolet behaviour of fermion fields depends critically on the
dimension of the space. We assume, to minimize infrared troubles, that
the latter is compact. The simplest case corresponds to $1+1$ field
theory, with chiral fermions living on an ${\Bbb R}\times{\Bbb T}^1$
spacetime. Let $F$ be the ordinary Dirac phase operator defining the
Wick ordering prescription and let $X$ denote a gauge transformation.
Choose the associated Fock space representation of the CAR algebra.
Then $[F,X]$ is Hilbert--Schmidt, and so the loop groups of arbitrary
Yang--Mills theories are contained in the group of Bogoliubov
transformations, and the ordering prescription by itself regularizes
the theory. This fact is behind the success of second quantization
methods in the construction of representations of the Virasoro and
Kac--Moody algebras~\cite{PressleyS}, and partly behind the development
of conformal field theory~\cite{Fuchs}. In the next odd case, $1+3$
field theory, which mainly concerns us, the ultraviolet behaviour, as
gauged by the summability of $[F,X]$, is much worse, and extra
renormalizations are needed in order to regularize the theory.

A long-standing hope, now amenable to rigorous scrutiny, is that
giving up locality, one of the basic tenets of rigorous quantum field
theory~\cite{HaagKastler} ---and indeed, one of the main selling points
by the forefathers~\cite{Jordan}--- will be rewarded with a better
ultraviolet behaviour. After all, noncommutative manifolds ---with NC
tori with irrational parameters as a case in point--- usually are much
more disconnected that ordinary ones. We shall see that this hope is
not borne out.

The content of the paper is as follows. First, we describe a general
framework for fermion fields on noncommutative spaces, in the presence
of background fields treated adynamically. For that, we recall in
Section~2 Connes' axioms for noncommutative fermionic single-particle
spaces. We check the axioms and exhibit the spin structure effecting
the neutrino paradigm~\cite{Marshak} over the noncommutative $3$-torus.
In Section~3 we discuss the Fredholm module structure. With the
(general) Dirac phase operator $F$ in hand, we proceed to second
quantization. The space of spinors on the algebra is an infinite
dimensional linear spinor space; we recall in Section~4 the definition
of the spin representation for its orthogonal group, whose
infinitesimal version yields the quantization prescription for the
currents. The construction of the scattering matrix is left for
another day, our main purpose here being to show how simple
noncommutative quantum field theory really is ---and why it belongs in
the toolkit of every theorist. We then examine the issue of the
ultraviolet behaviour by means of our example. In Section~5 we see by
direct computation that, as ``measured'' by the stick considered in
this paper, the ultraviolet behaviour of the theory is the same as for
a commutative torus. Finally, in Section~6 we discuss why such
behaviour of NC tori should be expected, on general grounds, on any
noncommutative manifold. This is related to some of the deeper issues
in noncommutative geometry.

The next logical step is to quantize bosonic actions for
(noncommutative) gauge fields, perhaps in the presence of external
currents. Then it would be time to tackle the full-blown
renormalization theory for nonlinear field configurations.

\section{First quantization on noncommutative tori}

Our method of work in this section is the following: each time that we
introduce basic data or axioms, we illustrate/comment on the
commutative case and verify them for the noncommutative $3$-torus. We
rely heavily on our Ref.~\onlinecite{Portia}. We begin, then, by
making  explicit the objects of a spin geometry
$({\cal A},{\cal H},D,J)$ for $3$-tori.

Let $\theta$ be a real skewsymmetric $n \times n$ matrix with
entries~$\theta_{jk}$. The $C^*$-algebra determined by~$n$ unitary
generators, with the relations
\[
u_k u_j = e^{2\pi i\theta_{jk}} u_j u_k,
\]
is called the $n$-torus algebra $A_\theta$. We focus on the $n = 3$
case with irrational entries $\theta_{jk}$. It is very convenient
---and suggested by consideration of the Weyl algebra--- to introduce
the unitary elements
\[
u^r := \exp\{\pi i(r_1\theta_{12}r_2 + r_1\theta_{13}r_3
             + r_2\theta_{23}r_3)\} \, u_1^{r_1} u_2^{r_2} u_3^{r_3}
\]
for each $r \in {\Bbb Z}^3$; the coefficient is chosen so that
$(u^r)^* = u^{-r}$ in all cases. They obey the product rule:
\[
u^r\,u^s = \lambda(r,s)\, u^{r+s},  \qquad
  \lambda(r,s) := \exp\{-\pi i\, r_j \theta_{jk} s_k\}.
\]
The noncommutative torus proper ${\cal A}_\theta := {\Bbb T}_\theta^3$
is the dense subalgebra of $A_\theta$ of ``noncommutative Fourier
series'':
\[
{\Bbb T}_\theta^3
 := \{\,a = a_r\, u^r : \{a_r\} \in {\cal S}({\Bbb Z}^3)\,\},
\]
where the coefficients belong to the space ${\cal S}({\Bbb Z}^3)$ of
rapidly decreasing sequences, namely, those for which
$(1 + |r|^2)^k \,|a_r|^2$ is bounded for all $k = 1,2,3,\dots$. In the
commutative case $\theta = 0$, we then have
${\Bbb T}_0^3 \simeq C^\infty({\Bbb T}^3)$.

On each torus algebra $A_\theta$ there is a faithful tracial
state~$\tau$, given by $\tau(a_r u^r) := a_0$. If $\theta$ is
irrational, the tracial state $\tau$ on $A_\theta$ is unique. Any
state on a $C^*$-algebra $A_\theta$ gives rise to a Hilbert space by
the well-known Gelfand--Na\u{\i}mark--Segal construction. So we
introduce the auxiliary Hilbert space ${\cal H}_0$ given as the
completion of the vector space $A_\theta$ in the Hilbert norm
\[
\|a\|_2 := \sqrt{\tau(a^*a)}.
\]
Since $\tau$ is a faithful state, the obvious map
$A_\theta \to {\cal H}_0$ is injective; we shall denote
by~$\underline{a}$ the vector in ${\cal H}_0$ corresponding to
$a \in A_\theta$. The GNS representation of $A_\theta$ is just
\[
\pi(a)\colon \underline{b} \mapsto \underline{ab}.
\]

We now look for the involution $J$. The obvious candidate to try is
\[
J_0(\underline{a}) := \underline{a^*}.
\]
(This is in fact the Tomita involution~\cite{KadisonR} determined by
the cyclic and separating vector~$\underline{1}$ for the algebra
$A_\theta$.) Notice, however, that $J_0^2 = +1$, whereas we require
$J^2 = -1$ in three dimensions (see Axiom~1 below). A simple device
allows us to modify the sign: we double the GNS Hilbert space by
taking ${\cal H} := {\cal H}_0 \oplus {\cal H}_0$ and define
\[
J := \pmatrix{0 & -J_0 \cr J_0 & 0 \cr}.
\]
The torus algebra acts on~$\cal H$ by the representation
\[
\pi(a) := \pmatrix{a & 0 \cr 0 & a \cr}.
\]
When $a \in {\cal A}$ and $\xi \in {\cal H}$, we shall usually write
$a\xi := \pi(a)\xi$. The vectors $\psi_m = \underline{u^m} \oplus 0$
and $\psi'_m = 0 \oplus \underline{u^m}$, for $m \in {\Bbb Z}^3$, form
a convenient orthonormal basis of~${\cal H}$.

Finally, we produce $D$. Let us consider the usual Pauli matrices
$\sigma_j$, and the derivations $\delta_1$, $\delta_2$, $\delta_3$
given by
\[
\delta_j(a_r u^r) := 2\pi i\,r_j\, a_r u^r,  \qquad  (j = 1,2,3).
\]
We define
\[
D := -i(\sigma_1\,\delta_1 + \sigma_2\,\delta_2 + \sigma_3\,\delta_3)
   = -i\pmatrix{\delta_3 & \delta_1 - i\delta_2  \cr
                \delta_1 + i\delta_2 & -\delta_3 \cr}
\]
Then $D^2 = - (\sigma\cdot\delta)^2
 = (-\delta_1^2 - \delta_2^2 - \delta_3^2)
   \pmatrix{1 & 0 \cr 0 & 1 \cr}$. This operator is diagonalized by the
orthonormal basis $\{\psi_m, \psi'_m\}$ of~${\cal H}$, with eigenvalues
$4\pi^2 |m|^2$. Using this basis we may express $D$, its absolute
value $|D|$ and the phase operator $F := D|D|^{-1}$ as (matrix)
multiplication operators in the index~$m$:
\[
D = 2\pi\, m\cdot\sigma,  \qquad  |D| = 2\pi\,|m|,  \qquad
F = \frac{m\cdot\sigma}{|m|}.
\]
The eigenvalues are then the same as for the ordinary Dirac operator
on the ordinary torus (with untwisted boundary conditions). One can
even introduce ``coherent spin states'' as eigenvectors of $F$: our
geometry looks like, and is, a spin one-half system on the NC tori.

Before introducing the further relations and properties that the
objects of a spin geometry, in particular for ${\Bbb T}^3_\theta$,
must satisfy, we make some precisions of a general nature on the data
themselves.

(1) A pre-$C^*$-algebra $\cal A$ is a dense involutive subalgebra of a
$C^*$-algebra $A$ that is stable under the holomorphic functional
calculus; or, more simply, such that the inverse (in~$A$) of any
invertible element of~$\cal A$ lies also in~$\cal A$. This happens,
for instance, whenever $\cal A$ is the smooth domain of a Lie algebra
of densely defined derivations of~$A$, since
$\delta(a^{-1}) = - a^{-1}\,\delta(a)\,a^{-1}$ for any derivation. The
major consequence of stability under the holomorphic functional
calculus is that the $K$-theories of~$\cal A$ and of~$A$ are the
same~\cite{Bost}.

For the algebra $A_\theta$, the common domain of the powers
$\delta_j^k$ of the commuting derivations $\delta_1,\delta_2,\delta_3$
is precisely the subalgebra ${\Bbb T}_\theta^3$; it is clear then that
${\Bbb T}_\theta^3$ is a pre-$C^*$-algebra.

\vspace{3pt}

(2) That $(D - \lambda)^{-1}$ is compact implies that $D$ has a
discrete spectrum of eigenvalues of finite multiplicity. This is
assured for the Dirac operator on a compact spin manifold. In most
circumstances the finite-dimensional kernel of $D$ is of no
consequence, and we have felt free to use the notation $D^{-1}$ when
convenient.

In the noncommutative case, we shall also refer to $D$ as the Dirac
operator. Connes' axioms are reorganized as follows: three with
algebraic flavour, three ``analytical'' axioms and lastly a
``topological'' one. (Such labels are a bit deceptive, of course.)

\vspace{6pt}

Axiom 1 (Reality):
The antilinear isometry $J \colon {\cal H} \to {\cal H}$ is such that
the representation given by $\pi^0(b) := J\pi(b^*)J^\dagger$ commutes
with $\pi({\cal A})$. Moreover the isometry satisfies
\[
J^2 = \pm 1,  \qquad  JD = \pm DJ,
\]
where the signs are precisely given by the following table:

\begin{center}
\begin{tabular}{|c|c|c|c|c|}
\hline
\rule[-5pt]{0pt}{16pt} $n \bmod 8$   & $1$ & $3$ & $5$ & $7$  \\
\hline
\rule[-5pt]{0pt}{16pt} $J^2 = \pm 1$ & $+$ & $-$ & $-$ & $+$  \\
\rule[-5pt]{0pt}{16pt} $JD = \pm DJ$ & $-$ & $+$ & $-$ & $+$  \\
\hline
\end{tabular}
\end{center}

This table arises from the structure of real Clifford algebra
representations that underlie $KR$-theory. It is well known that, in
the commutative case of Riemannian spin manifolds, one can find
conjugation operators $J$ on spinors that satisfy these sign rules.

In the noncommutative case, the antilinear operator $J$ comes from the
Tomita involution on a Hilbert space: $\pi^0$ is a representation of
the opposite algebra ${\cal A}^0$, consisting of elements
$\{\,a^0 : a \in {\cal A}\,\}$ with product $a^0b^0 = (ba)^0$ ---we
can write $b^0 = J b^* J^\dagger$. We have thus required that the
representations $\pi$ and~$\pi^0$ commute. When $\cal A$ is
commutative, we may also require $J\pi(b^*)J^\dagger = \pi(b)$,
whereupon the commutation of representations is automatic.

For $3$-tori, the opposite algebra $A_\theta^0$ is just $A_{-\theta}$,
and the commuting representation of $A_{-\theta}$ on~${\cal H}_0$ is
given by right multiplication by elements of~$A_\theta$:
\[
a^0\,\underline{b} = J_0 a^* J_0^\dagger\,\underline{b}
 = J_0\,\underline{a^*b^*} = \underline{ba}.
\]
From that, verification of the reality axiom is immediate.

\vspace{6pt}

Axiom 2 (First-order property):
For all $a,b \in {\cal A}$, the following commutation relation
moreover holds:
\[
[[D, a], J b^* J^\dagger] = 0.
\]

That can be rewritten as $[[D,\pi(a)],\pi^0(b)] = 0$. In view of this
condition, the bimodule over~$\cal A$ given by
$C_n({\cal A},{\cal A} \otimes {\cal A}^0)
 := ({\cal A} \otimes {\cal A}^0) \otimes {\cal A}^{\otimes n}$ is
represented by operators on~$\cal H$:
\begin{eqnarray*}
&&\pi_D((a \otimes b^0) \otimes a_1 \otimes a_2
  \otimes\cdots\otimes a_n)
\\
&&\quad := \pi(a) \pi^0(b) \,[D,\pi(a_1)] \,[D,\pi(a_2)] \dots
            [D,\pi(a_n)].
\end{eqnarray*}
The elements of $C_n({\cal A},{\cal A} \otimes {\cal A}^0)$ are called
Hochschild $n$-chains with coefficients in
${\cal A} \otimes {\cal A}^0$.

In the commutative case, we may replace ${\cal A} \otimes {\cal A}^0$
simply by $\cal A$, and the axiom expresses the fact that the Dirac
operator $D\mkern-11.5mu/\,$ is a first-order differential operator.

The first-order axiom for our NC $3$-torus geometry can be readily
checked, using the fact that $D$ comes from a derivation of the
algebra.

\vspace{6pt}

Axiom 3 (Orientability):
There exists a {\it Hochschild cycle\/}
$\bbox{c} \in Z_n({\cal A},{\cal A} \otimes {\cal A}^0)$ whose
representative on~$\cal H$ fulfills
\[
\pi_D(\bbox{c}) = 1.
\]
We say that the Hochschild $n$-chain $\bbox{c}$ is a cycle when its
boundary is zero, where the Hochschild boundary operator for $n = 3$
is
\begin{eqnarray*}
&&b(m_0 \otimes a_1 \otimes a_2 \otimes a_3)
\\
&&\qquad :=        m_0 a_1 \otimes a_2 \otimes a_3
                 - m_0 \otimes a_1 a_2 \otimes a_3
\\
&&\qquad\qquad{} + m_0 \otimes a_1 \otimes a_2 a_3
                 - a_3 m_0 \otimes a_1 \otimes a_2,
\end{eqnarray*}
for $m_0 \in {\cal A} \otimes {\cal A}^0$; and similarly for
other~$n$. Then $b^2 = 0$, making
$C_\bullet({\cal A},{\cal A} \otimes {\cal A}^0)$ a chain complex.

The Hochschild cycle $\bbox{c}$ is the algebraic equivalent of a
volume form, on a noncommutative manifold. Indeed, in the commutative
case, a volume form is a sum of terms
$a_0\,da_1 \wedge\cdots\wedge da_n$, which we represent by an
antisymmetric sum:
\[
\bbox{c}' := \sum_\sigma (-)^\sigma a_0 \otimes a_{\sigma(1)}
               \otimes\cdots\otimes a_{\sigma(n)}
\]
in ${\cal A}^{\otimes(n+1)} = C_n({\cal A},{\cal A})$. Then
$b\bbox{c}' = 0$ by cancellation since $\cal A$ is commutative. When
${\cal A} = C^\infty(M)$, chains are represented by Clifford products:
$\pi_{D\mkern-11.5mu/\,}(a_0 \otimes a_1 \otimes\cdots\otimes a_n)
 = a_0\,\gamma(da_1)\dots \gamma(da_n)$, with
$\gamma(da) := \gamma^j \,\partial_j a$, where the $\gamma^j$ are
essentially the Dirac matrices.

For our $3$-torus geometries, consider the Hochschild chain:
\[
\bbox{c} := \frac{1}{6(2\pi i)^3}\, \epsilon^{ijk}
         u_i^{-1}u_j^{-1}u_k^{-1} \otimes u_k \otimes u_j \otimes u_i.
\]
(The first tensor factor can lie in~${\cal A}$, since
${\cal A} \simeq {\cal A} \otimes 1^0
 \subset {\cal A} \otimes {\cal A}^0$.) We check that this $\bbox{c}$
is a Hochschild $3$-cycle on any ${\Bbb T}_\theta^3$. In fact,
\begin{eqnarray*}
\lefteqn{6(2\pi i)^3 \, b\bbox{c}}
\\
&=& \epsilon^{ijk}\, (u_i^{-1}u_j^{-1} \otimes u_j \otimes u_i
           - u_i^{-1}u_j^{-1}u_k^{-1} \otimes u_k u_j \otimes u_i
\\
&&\qquad{} + u_i^{-1}u_j^{-1}u_k^{-1} \otimes u_k \otimes u_j u_i
           - u_j^{-1}u_k^{-1} \otimes u_k \otimes u_j)
\end{eqnarray*}
and the first and fourth terms cancel after cyclic permutation of the
indices. Therefore
\begin{eqnarray*}
6(2\pi i)^3 \, b\bbox{c}
&=& \epsilon^{ijk}\,
   (u_i^{-1} u_j^{-1} u_k^{-1} \otimes u_k \otimes u_j u_i
\\
&&\qquad{} - u_i^{-1} u_j^{-1} u_k^{-1} \otimes u_k u_j \otimes u_i).
\end{eqnarray*}
The remaining term $\epsilon^{ijk}\, u_i^{-1} u_j^{-1} u_k^{-1}
 \otimes u_k \otimes u_j u_i$ vanishes by antisymmetrization, since
the commutation relations imply
\[
u_i^{-1} u_j^{-1} u_k^{-1} \otimes u_k \otimes u_j u_i
 = u_j^{-1} u_i^{-1} u_k^{-1} \otimes u_k \otimes u_i u_j.
\]
Likewise the second remaining term vanishes by antisymmetrization.

The representative on~$\cal H$ given by the geometry is the identity;
in effect, $[D,u_j] = -i\sigma_j\,\delta_j(u_j) = 2\pi\sigma_j\,u_j$,
and therefore
\[
\pi_D(\bbox{c}) = \frac{(2\pi)^3}{6(2\pi i)^3}
  \epsilon^{ijk}\,\sigma_k \sigma_j \sigma_i
 = \frac{-6i\,(2\pi)^3}{6(2\pi i)^3} = 1.
\]

\vspace{6pt}

Axiom 4 (Classical Dimension):
There is an {\it integer\/} $n$, the classical dimension of the spin
geometry, for which the singular values of~$|D|^{-n}$ form a
logarithmically divergent series. The coefficient of logarithmic
divergence will be denoted by $\ncint ds^n$.

In our case, since $D^2 = 4\pi^2|m|^2$ on a $2$-dimensional eigenspace
for each~$m$, we see that
\begin{eqnarray*}
\ncint ds^n
&=& 2 \lim_{R\to\infty} \frac{1}{3\log R}
                \sum_{1\leq |m|\leq R} (2\pi|m|)^{-n}
\\
&=& 2 \lim_{R\to\infty} \frac{1}{3\log R}
     \int_1^R \frac{4\pi r^2\,dr}{(2\pi r)^n},
\end{eqnarray*}
which is zero for $n > 3$, diverges for $n < 3$ and is positive finite
(equal to $(3\pi^2)^{-1}$) for $n = 3$; so indeed the dimension is~$3$.

Once we know what the correct dimension for a noncommutative manifold
$\cal A$ is, we write $\ncint a\,ds^n$ for the coefficient of
logarithmic divergence of $a|D|^{-n}$, that exists for $a \in A$. In
the commutative case, denoting by $\mu$ the canonical measure, Connes'
trace theorem (see Section~6) shows that we have
$\ncint a\,ds^n = C_n \int a(x) \,d\mu(x)$, with $C_n$ a normalization
factor. In dimension~$3$, the normalization factor is precisely
$1/3\pi^2$~\cite{Sirius}.

Note that we have for $3$-tori: $\ncint u^r\,ds^3 = 0$ unless $r = 0$.
This can be proved, for instance, by using the zeta-function
recipe~\cite{Kassel} for the computation of the noncommutative
integral: $\ncint u^r\,ds^3
 = \mathop{\rm Res}_{s=1} \mathop{\rm Tr}(u^r |D|^{-3s}) = 0$ since,
for any $r \neq 0$ and $s > 1$, $u^r|D|^{-3s}$ is a traceclass
operator with an off-diagonal matrix.

\vspace{6pt}

Axiom 5 (Regularity):
For any $a \in {\cal A}$, the operator $[D,a]$ is bounded on~$\cal H$,
and both $a$ and $[D,a]$ belong to the domain of smoothness
$\bigcap_{k=1}^\infty \mathop{\rm Dom}(\delta^k)$ of the derivation
$\delta$ on~$L({\cal H})$ given by $\delta(T) := [|D|,T]$.

The regularity axiom has far-reaching implications. As shown by
Cipriani {\it et~al}~\cite{CiprianiGS}, it implies, in particular, that
$\ncint$ is a trace on the algebra $\cal A$; i.e.,
$\ncint ab\,ds^n = \ncint ba\,ds^n$ for all $a, b \in {\cal A}$. This
finite trace on~$\cal A$ extends to a finite normal trace on the
von~Neumann algebra ${\cal A}''$ generated by~$\cal A$; therefore
${\cal A}''$ can only have components of types ${\rm I}_n$ and
${\rm II}_1$~\cite{KadisonR}.

In the commutative case, where $[D\mkern-11.5mu/\,,a] = \gamma(da)$,
this axiom amounts to saying that $a$ has derivatives of all orders,
i.e., that ${\cal A} \subseteq C^\infty(M)$. This is proved with the
pseudodifferential calculus. Consequently, all multiplication
operators in $\bigcap_{k=1}^\infty \mathop{\rm Dom}(\delta^k)$ are
multiplications by smooth functions.

Verification of the regularity axiom for our noncommutative torus is
straightforward.

\vspace{6pt}

Axiom 6 (Finiteness):

Denote by $\langle\cdot\mathbin|\cdot\rangle$ the inner product on
$\cal H$. The space of smooth vectors
${\cal H}_\infty := \bigcap_{k=1}^\infty \mathop{\rm Dom}(D^k)$ is a
finite projective left $\cal A$-module with a Hermitian structure
$(\cdot\mathbin|\cdot)$ defined by
\[
\ncint (\xi\mathbin|\eta)\,ds^n = C_n\,\langle\xi\mathbin|\eta\rangle.
\]

The axiom assumes the trace property for the noncommutative integral,
as we see from the following manipulation:
\begin{eqnarray*}
\ncint a\,(\xi\mathbin|\eta) \,ds^n
&=& \ncint (\xi\mathbin| a\eta) \,ds^n
 = C_n\, \langle\xi\mathbin| a\eta\rangle
\\
= C_n\, \langle a^*\xi\mathbin|\eta\rangle
&=& \ncint (a^*\xi\mathbin|\eta) \,ds^n
 = \ncint (\xi\mathbin|\eta)\,a \,ds^n.
\end{eqnarray*}
In the commutative case, Connes's trace theorem shows that
$(\xi\mathbin|\eta)$ is just the hermitian product of spinors given by
the metric on the spinor bundle. For our $3$-torus, plainly
${\cal H}_\infty = {\Bbb T}_\theta^3 \oplus {\Bbb T}_\theta^3$ is a
projective (indeed, free) left module over~${\Bbb T}_\theta^3$, and
the hermitian structure is also manifest.

\vspace{6pt}

Axiom 7 (P-duality):
The Fredholm index of the operator $D$ yields a nondegenerate
intersection form on the $K$-theory of the algebra
${\cal A} \otimes {\cal A}^0$.

We shall not discuss it here for the NC $3$-torus, except to say that
the $K$-theory groups of the $3$-tori are
$K_j({\Bbb T}_\theta^3) \simeq {\Bbb Z}^4$ for $j = 0,1$, and
all~$\theta$.

\vspace{6pt}

If we add an ``Axiom 0'', establishing that $\cal A$ is the
commutative algebra $C^\infty(M)$ of smooth functions on a compact
manifold $M$, then $M$ is spin, and there is a distinguished
representation of the geometry for which $\pi$ is unitarily equivalent
to the representation of $\cal A$ by multiplication operators on the
canonical spinor space, and $D$ to the canonical Dirac
operator~$D\mkern-11.5mu/\,$~\cite{ConnesAxioms}. Also, $C^\infty(M)$
is Morita equivalent to the Clifford algebra over $M$~\cite{Plymen}.

Of course, Axiom~7 is then redundant. It is to be hoped that the same
conclusions may be obtained by just stipulating commutativity of the
algebra; but we know no proof of that yet.

At any rate, it transpires that the previous axioms constitute an
appropriate description of noncommutative spin manifolds. To be sure,
much work remains to be done: we do not have classification results.

In general, the fermions will be coupled to a given ``external''
Yang--Mills configuration, that may be time-dependent, but whose
dynamics is not involved in the problem. For the commutative geometry
$(C^\infty(M), L^2(M,S), D\mkern-11.5mu/\,, J)$, we may have a
nonabelian gauge theory, formulated on a Hermitian $G$-vector bundle
$E$ over $M$. The Dirac operator then acts on the Hilbert space
$L^2(M, S \otimes E)$. Gauge transformations are elements of the group
$C^\infty(\mathop{\rm Aut} E)$~\cite{FreedU}. Pointwise multiplication
gives the representation of $C^\infty(\mathop{\rm Aut} E)$ on the
Hilbert space. When $E$ is trivial,
$C^\infty(\mathop{\rm Aut} E) \simeq \mathop{\rm Map}(M,G)$.
Infinitesimal gauge transformations are accordingly defined. Gauge
potentials, in the commutative case, are $E$-valued $1$-forms on~$M$,
represented on spinor space as Clifford multiplication operators. In
the noncommutative case, vector bundles are translated into finitely
generated projective (right) modules over the algebra~${\cal A}$. The
vector bundles over noncommutative $n$-tori have been all
constructed~\cite{RieffelP} and partially classified up to Morita
equivalence~\cite{RieffelS}, and the corresponding gauge
transformations are easily determined. Gauge potentials can also be
translated to the noncommutative case~\cite{Portia}. In what follows,
we leave aside all geometrical complications extraneous to the
analytical problem at hand.

\section{A Fredholm module interlude}

A {\it cycle\/} is a complex graded associative algebra
$\Omega^\bullet = \bigoplus_{k=0}^\infty \Omega^k$, endowed with a
differential $d\colon \Omega^\bullet \to \Omega^\bullet$, i.e., a
linear map of degree~$+1$ such that $d^2 = 0$ and
\[
d(\omega_k\omega_l) = (d\omega_k)\,\omega_l + (-)^k \omega_k\,d\omega_l
\]
when $\omega_k$, $\omega_l$ are homogeneous elements of respective
degrees $k,l$; and with an integral $\int$, namely, a linear map
$\int\colon \Omega^\bullet \to {\Bbb C}$ such that
\[
\int \omega_k\omega_l = (-)^{kl}\int \omega_l\omega_k
 \mbox{ \ and } \int\,d\omega = 0 \mbox{ for any }
  \omega \in \Omega^\bullet.
\]
We refer to the last property as closedness of the integral. A cycle
over an algebra ${\cal A}$ is a cycle $(\Omega^\bullet, d, \int)$
together with a homomorphism from ${\cal A}$ to~$\Omega^0$. The
simplest examples are afforded by de~Rham complexes.

A truly interesting class of examples comes from {\it Fredholm
modules\/} over a given algebra $\cal A$. An odd Fredholm module over
$\cal A$ is given by an involutive representation $\pi$ of $\cal A$ on
a Hilbert space $\cal H$ and a symmetry (selfadjoint unitary operator)
$F$ such that $[F,\pi(a)]$ is a compact operator for all
$a \in {\cal A}$. Let ${\cal H}^\pm$ denote the eigenspaces for the
$\pm 1$ eigenvalues of $F$. Then we may write any operator $T$ as
\[
T = \pmatrix{\alpha & \beta \cr \gamma & \delta},
\]
where $\alpha\colon {\cal H}^+ \to {\cal H}^+$,
$\beta\colon {\cal H}^- \to {\cal H}^+$ and so on. For a given $F$,
$T$ is thus decomposed into ``linear'' and ``antilinear'' parts:
\begin{eqnarray*}
T = T_+ + T_-
&:=& \case{1}{2} (T + FTF) + \case{1}{2} (T - FTF)
\\
&=& \pmatrix{\alpha & 0 \cr 0 & \delta \cr}
    + \pmatrix{0 & \beta \cr \gamma & 0 \cr}.
\end{eqnarray*}

To define an integral, let us postulate a summability condition on the
algebra: for all $a \in {\cal A}$ and for some chosen nonnegative
integer $n$, we assume that $a_-$ belongs to the Schatten class
${\cal L}^{n+1}({\cal H})$. The graded differential algebra structure
is introduced as follows: define $\Omega^k({\cal A})$ as the space
spanned by forms $a_0\,{\rm d}a_1\dots {\rm d}a_k$ with
$a_0,a_1,\dots,a_k \in A$, where ${\rm d}a := [F,a]$. The algebra
multiplication is the operator product. Given an operator $T$ on
$\cal H$, we introduce its {\it conditional\/} trace:
\[
\mathop{\rm Tr}\nolimits_C T := \mathop{\rm Tr} T_+.
\]
Note that $\mathop{\rm Tr}\nolimits_C(AB) = \mathop{\rm Tr}\nolimits_C(BA)$ when both
sides make sense, and that $\mathop{\rm Tr}\nolimits_C T := \mathop{\rm Tr} T$,
if $T \in {\cal L}^1$, by cyclicity of the trace. Assuming that $n$ is
odd, one has $(\omega_n)_+ \in {\cal L}^1$ \cite{ConnesIHES,Book}.
Therefore, it makes sense to define the integral by
\[
\int \omega_n := \mathop{\rm Tr}\nolimits_C \omega_n
 = \case{1}{2} \mathop{\rm Tr}\,F\,{\rm d}\omega_n.
\]
We shall then say that (the cycle associated to) the Fredholm module
has dimension~$n$. The {\it Chern character\/} of that cycle is
defined to be the $(n+1)$-linear functional on~${\cal A}$ given by
\[
\tau(a_0,a_1,\dots,a_n)
 := \mathop{\rm Tr}\nolimits_C (a_0\,{\rm d}a_1\,{\rm d}a_2 \dots {\rm d}a_n).
\]
We have $b\tau = 0$, since
\begin{eqnarray*}
\lefteqn{\mathop{\rm Tr}\nolimits_C
    ((a_0a_1\, {\rm d}a_2 \dots {\rm d}a_{n+1})}
\\
&+& \sum_{i=1}^n (-)^i \mathop{\rm Tr}\nolimits_C
   (a_0\,{\rm d}a_1 \dots ({\rm d}a_i\, a_{i+1} + a_i\,{\rm d}a_{i+1})
    \dots {\rm d}a_{n+1})
\\
&&\qquad + (-)^{n+1}
     \mathop{\rm Tr}\nolimits_C (a_{n+1}a_0\,{\rm d}a_1 \dots {\rm d}a_n)
\\
&=& (-)^n \mathop{\rm Tr}\nolimits_C
    ((a_0\,{\rm d}a_1\dots{\rm d}a_n) a_{n+1})
\\
&&\qquad + (-)^{n+1} \mathop{\rm Tr}\nolimits_C
    (a_{n+1}a_0\,{\rm d}a_1 \dots {\rm d}a_n) = 0,
\end{eqnarray*}
by telescoping; the last equality is just the trace property
$\int a \omega = \int \omega a$ for $a \in \Omega^0$,
$\omega \in \Omega^n$. Thus $\tau$ is an $n$-cocycle. Moreover, $\tau$
is {\it cyclic\/}:
\begin{eqnarray*}
\lefteqn{\tau(a_0,a_1,\dots,a_n)
 = (-)^{n-1} \mathop{\rm Tr}\nolimits_C
   ({\rm d}a_2\dots{\rm d}a_n\,a_0\,{\rm d}a_1)}
\\
&=& (-)^n \mathop{\rm Tr}\nolimits_C({\rm d}a_2\dots{\rm d}a_n\,{\rm
d}a_0\,a_1)
\\
&=& (-)^n \mathop{\rm Tr}\nolimits_C
   (a_1\,{\rm d}a_2\dots{\rm d}a_n\,{\rm d}a_0)
 = (-)^n \tau(a_1,\dots,a_n,a_0),
\end{eqnarray*}
where we have used that
${\rm d} a_0\, a_1 + a_0\,{\rm d} a_1 = {\rm d}(a_0a_1)$ and the
closedness of~$\mathop{\rm Tr}\nolimits_C$.

Given a Dirac operator $D$ on a spin geometry of dimension~$n$ (e.g.,
an $n$-torus), there is a God-given Fredholm module coming from the
phase operator $D/|D|$. The minimal integer for which the character
exists, for this Fredholm module structure, we call the ``quantum
dimension'' of the spin space. Note that (non)commutativity of
$\cal A$ does not play any r\^ole in the foregoing.

\section{Second quantization}

We now review the algebraic machinery of canonical quantization, and
investigate its general properties of application. It is important to
realize that the basic ingredient of the construction is just a real
Hilbert space. So suppose an infinite-dimensional real vector space
$V$ and a symmetric bilinear form~$d$ are given, the metric space
$(V,d)$ being complete. The first object in quantization is the field
algebra over the space $(V,d)$, which is just the complexified
Clifford algebra
${\frak A}(V) := \mathop{\rm C\ell}(V,d) \otimes {\Bbb C}$, complete
in the (inductive limit) $C^*$-norm~\cite{Emch}. The {\it fermion
field\/} is a linear map $B\colon V \to {\frak A}(V)$ satisfying
$[B(v), B(v')]_+ = 2\,d(v,v')$ for all $v,v' \in V$. Any two
$C^*$-algebras generated by two sets of operators obeying the same
rules are isomorphic~\cite{ArakiB}.

The orthogonal group $O(V)$ is $\{\,g\in GL_{\Bbb R}(V)
 : d(gu,gv) = d(u,v) \hbox{ for all } u,v\in V\,\}$. A {\it complex
structure\/}~$K$ is an orthogonal operator on~$V$ satisfying
$K^2 = -I$. Now, introducing the rule
$(\alpha + i\beta)v := \alpha v + \beta Kv$ for $\alpha,\beta$ real,
the hermitian form
\[
\langle u\mathbin|v\rangle_{\scriptscriptstyle K} := d(u,v) + id(Ku,v)
\]
makes $(V,d,K)$ a complex Hilbert space. Once a particular complex
structure $K$ has been selected, one can decompose elements of $O(V)$
as $g = p_g + q_g$ where $p_g$, $q_g$ are its linear and antilinear
parts: $p_g := \case{1}{2}(g - KgK)$, $q_g := \case{1}{2}(g + KgK)$.
The ``restricted orthogonal group'' $O_K(V)$ is the subgroup of~$O(V)$
consisting of those $g$ for which $q_g$ is Hilbert--Schmidt.

One can construct a faithful irreducible representation $\pi_K$ of
${\frak A}(V)$ by the GNS construction with respect to the ``Fock
state'' $\omega_K$ determined by $\omega_K(B(u)B(v))
 := \langle u\mathbin|v\rangle_{\scriptscriptstyle K}$; this is the
standard representation on the fermion Fock space ${\cal F}_K(V)$,
with vacuum $\Omega$, in which the creation and annihilation operators
are defined as real-linear operators:
\[
a_K^\dagger(v) := \pi_KB(P_Kv),  \qquad  a_K(v) := \pi_KB(P_{-K}v),
\]
where $P_K := \case{1}{2}(I - iK)$.

For $g$ orthogonal, the map $w \mapsto B(gw)$ extends to a
$*$-automorphism of the CAR algebra~${\frak A}(V)$. We then ask when
these two quantizations are unitarily equivalent, i.e., whether this
$*$-automorphism is unitarily implementable on~${\cal F}_K(V)$. For a
given $g\in O(V)$, we seek a unitary operator $\mu(g)$
on~${\cal F}_K(V)$ so that
\[
\mu(g) B(v) = B(gv) \mu(g),  \qquad\mbox{for all}\quad  v \in V.
\]

The complex structure $K$ is transformed to $gKg^{-1}$; the creation
and annihiliation operators undergo a {\it Bogoliubov
transformation\/}:
\begin{eqnarray*}
a_{gKg^{-1}}^\dagger(gv) &=& a_K(q_gv) + a_K^\dagger(p_gv),
\\
a_{gKg^{-1}}(gv) &=& a_K(p_gv) + a_K^\dagger(q_gv).
\end{eqnarray*}
Were $\mu(g)$ to exist, we would then have
\begin{eqnarray*}
\mu(g) a_K^\dagger(v) &=& a_{gKg^{-1}}^\dagger(gv) \mu(g),
\\
\mu(g) a_K(v) &=& a_{gKg^{-1}}(gv) \mu(g).
\end{eqnarray*}
The out-vacuum $\mu(g)\Omega$ is annihilated by
$a_{gKg^{-1}}(gv)$, for all $v \in V$. From there the
Shale--Stinespring criterion~\cite{ShaleS} for implementability is
easily established: the operator $\mu(g)$ exists if and only if $g$
belongs to the restricted orthogonal group. Naturally, the map
$g \mapsto \mu(g)$ is only a projective representation of $O_K(V)$.
The explicit construction of $\mu$ was performed in our
Ref.~\cite{Rhea}, on which we mostly rely for this section.

The spin representation allows us to quantize all elements of the Lie
algebra ${\frak o}_{\scriptscriptstyle K}(V)$ of the group $O_K(V)$.
Define the {\it infinitesimal spin representation\/} $\dot\mu(X)$ of
$X \in {\frak o}_{\scriptscriptstyle K}(V)$ by:
\[
\dot\mu(X)\,\Psi
 := \frac{d}{dt}\biggr|_{t=0} e^{i\theta_X(t)} \mu(\exp tX) \,\Psi
\]
for $\Psi \in {\cal F}_K(V)$, where $\theta_X(t)$ is such that
$t \mapsto e^{i\theta_X(t)} \mu(\exp tX)$ is a homomorphism. The
vacuum expectation value of $\dot\mu(X)$ is
$\langle\Omega\mathbin|\dot\mu(X)\Omega\rangle = i\theta'_X(0)$. We
set $\theta'_X(0) = 0$ for
all~$X \in {\frak o}_{\scriptscriptstyle K}(V)$. The quantization rule
$X \mapsto \dot\mu(X)$ then is uniquely specified by the condition of
vanishing vacuum expectation values.

The fundamental property of the infinitesimal spin representation is
the commutation relations:
\[
[\dot\mu(X), B(v)] = B(Xv),
\]
an operator-valued equation valid on a dense domain
in~${\cal F}_K(V)$, that justifies the name ``currents'' for the
quantized observables. An easy computation~\cite{Rhea} gives
\[
[\dot\mu(X), \dot\mu(Y)] - \dot\mu([X,Y])
 = \case{i}{4} \mathop{\rm Tr}(K[K,X][K,Y])
\]
when $X,Y \in {\frak o}_{\scriptscriptstyle K}(V)$, for the Schwinger
terms.

We reexpress the quantization prescription in the language of creation
and annihilation operators. Given orthonormal bases $\{e_j\}$,
$\{f_k\}$ of $(V,d,K)$, the quadratic expressions:
\begin{eqnarray*}
a^\dagger T a^\dagger
&:=& \sum_{j,k} a_K^\dagger(f_k)
      \,\langle f_k\mathbin| Te_j\rangle \,a_K^\dagger(e_j),
\\
a T a
&:=& \sum_{j,k} a_K(e_j) \,\langle Te_j\mathbin|f_k\rangle \,a_K(f_k),
\\
a^\dagger C a
&:=& \sum_{j,k} a_K^\dagger(f_k)
      \,\langle f_k\mathbin| Ce_j\rangle \,a_K(e_j),
\end{eqnarray*}
are independent of the orthonormal bases used $T$ is antilinear and
skew, and $C$ is linear, as operators on~$V$. The series
$a^\dagger T a^\dagger$, $a T a$ are meaningful in Fock space if and
only if $T$ is Hilbert--Schmidt. If $C_X$ is the linear part of $X$
and $A_X$ the antilinear part, we thus get:
\begin{equation}
\dot\mu(X)
 = \case{1}{2}(a^\dagger A_X a^\dagger + 2 a^\dagger C_X a - a A_X a).
\label{eq:qn-rule}
\end{equation}

In most cases, including our neutrino fields over noncommutative tori,
$V$ is a complex Hilbert space to start with. The original complex
structure contains important physical information; but we have seen
that the first step of second quantization is to forego and replace it
with a new complex structure adapted to the dynamical problem at hand.
If $V$ has this additional structure, then unitary elements of
${\cal L}_{\Bbb C}(V)$ are obviously orthogonal; and selfadjoint
elements of ${\cal L}_{\Bbb C}(V)$ are of the form $iX$, with
$X \in {\frak o}_{\scriptscriptstyle K}(V)$. In this context, it is
plain that if $F$ is a symmetry define a Fredholm module, then $iF$
becomes a complex structure on the realification of~$\cal H$.

Suppose, moreover, that $F$ is the phase of the Dirac operator on a
spin geometry, commutative if one wishes, and let $M$ denote the
underlying manifold, with dimension~$n$. Then $F$ defines the very
complex structure we naturally use to quantize fermions over $M$: we
can think of the $F$-eigenspaces ${\cal H}_+$ and ${\cal H}_-$ as the
spaces of positive and negative energy solutions, respectively, of the
Dirac equation
\[
i\biggl( \frac{\partial}{\partial t} - D \biggr) \psi = 0,
\]
so the construction of the new Hilbert space with complex structure
$iF$ is equivalent to filling up the Dirac sea. In fact, $iD/|D|$ is
the {\it unique\/} complex structure for which $D$ becomes a positive
generator. Then the quantization prescription~(\ref{eq:qn-rule})
effects normal ordering; it is equivalent to the one defined
in~\cite{MLSaga} ---although our formalism is more general.

The outcome of the previous discussion is that an orthogonal operator
$O$ on the single-particle space can be second-quantized (by means of
the spin representation) to an operator on the Fock space associated
to the ``free'' evolution iff $[F,O]$ is Hilbert--Schmidt, and an
infinitesimally orthogonal operator $Z$ on the single-particle space
can be second-quantized (by means of the infinitesimal spin
representation) to an operator on the same Fock space iff $[F,Z]$ is
Hilbert--Schmidt. In the complex context, the orthogonal operator will
be actually in most cases unitary with respect to the original or
fiducial complex structure, and the infinitesimally orthogonal one
actually skewadjoint.

In the commutative case, if $g$ is a multiplication operator, we have
$[F,g] \in {\cal L}^{n+1}({\cal H})$. The proof relies on
pseudodifferential operators: if $T$ is pseudodifferential of order
$m < 0$, then it belongs to the Schatten class ${\cal L}^p$ for all
$p > -n/m$. This can be deduced from the Ces\`aro asymptotic
development of the spectral density of such operators~\cite{Odysseus}.
Now, $F$ and $g$ are of order~$0$, so $[F,g]$ is of order~$-1$.
Therefore $[F,g] \in {\cal L}^p({\cal H})$ for all $p > n$, in
particular for $p = n + 1$. As hinted at the end of Section~2, this
conclusion is not altered when $g$ is replaced by an element of a more
complicated projective module over $C^\infty(M)$, representing a gauge
theory on~$M$.

The Schatten class of $[F,g]$, thus the ``quantum dimension'',
measures the degree of ultraviolet divergence of the theory. We have
seen that, at least for commutative manifolds, the classical and
quantum dimensions coincide. For $1+1$ quantum field theory, the
character is identical to the Schwinger term; the Shale--Stinespring
criterion is satisfied for any~$g$, and so normal ordering is
sufficient to regularize the theory. In fact, it is even sufficient to
regularize the fully interacting gauged Wess--Zumino--Witten
model!~\cite{LangmannS}. This is not so for $1+3$ quantum field theory,
where the gauge transformations themselves cannot be unitarily
implemented in general.

\section{Quantum dimension = classical dimension for NC tori}

A gauge transformation for the trivial line bundle over
${\Bbb T}_\theta^3$ is just a unitary element $X$ of this algebra. For
irrational~$\theta$, ${\Bbb T}_\theta^3$ is a highly nonlocal algebra,
and one might expect that its quantum dimension would be less
than~$3$, namely, that typically $[F,X] \in {\cal L}^p$ for some
$p \leq 3$. But this is not the case: indeed, the nonlocality of the
irrational $3$-torus does nothing to improve that particular test of
ultraviolet behaviour.

We may write $X = a_r u^r$ with $\{a_r\} \in {\cal S}({\Bbb Z}^3)$;
then $X^* = \bar a_r u^{-r} = \bar a_{-s} u^s$ and
$X^*X = \lambda(m,r)\,\bar a_r a_{r+m} \,u^m$, so that $X$ is unitary
if and only if
\[
\mathop{\textstyle\sum}\nolimits_r |a_r|^2 = 1,  \qquad
\mathop{\textstyle\sum}\nolimits_r \lambda(m,r)\, \bar a_r a_{r+m}
 = 0  \mbox{ for } m \neq 0.
\]
The unitarity conditions in particular imply that a {\it finite\/} sum
$X = a_r u^r$ can be unitary only if it contains just one summand,
i.e., $X$ is a multiple of some~$u^r$.

We start from the computation carried out by Mickelsson and
Rajeev~\cite{MickelssonR} ten years ago for commutative tori. With
respect to the orthonormal basis
$\{\,\psi_n,\psi'_n : n \in {\Bbb Z}^3\,\}$ for~${\cal H}$, the matrix
entries of the operator $A = [F,X]$ are given by
\[
[F,X] \psi_n = \sum_r \lambda(r,n) a_r\, \biggl(
  \frac{(n+r)\cdot\sigma}{|n+r|} - \frac{n\cdot\sigma}{|n|} \biggr)
  \,\psi_{n+r},
\]
and similarly for the $\psi'_n$. To obtain the Schatten class of~$A$,
we must determine the finiteness of the $p$-norm
\[
\|A\|_p := \bigl( \mathop{\rm Tr}(A^*A)^{p/2} \bigr)^{1/p},
\]
that is in general hard to compute. A simpler alternative is to
calculate
\[
\snorm{A}_p
 := \biggl( \sum_n \|A\psi_n\|^p + \|A\psi'_n\|^p \biggr)^{1/p},
\]
or its analogue with any other orthonormal basis of~${\cal H}$.
However, these are not equivalent norms unless $p = 2$, {\it pace\/}
Ref.~\onlinecite{MickelssonR}. It is known~\cite{GohbergM} that
$\|A\|_p \leq \snorm{A}_p$ if $1 \leq p \leq 2$, whereas
$\snorm{A}_p \leq \|A\|_p$ if $p \geq 2$. Thus, in general, for
$p > 2$ the divergence of $\snorm{A}_p$ implies that
$A \notin {\cal L}^p$, but not conversely.

For the particular case $A = [F,u^r]$ this does not matter, since
$A^*A$ is diagonal in the chosen basis. Indeed,
\begin{eqnarray*}
\lefteqn{[F,u^r]^* [F,u^r] \psi_n}
\\
&=& \bar\lambda(r,n+r) \, \lambda(r,n) \biggl(
   \frac{(n+r)\cdot\sigma}{|n+r|} - \frac{n\cdot\sigma}{|n|} \biggr)^2
	\,\psi_n
\\
&=& 2 \biggl( 1 - \frac{(n+r)\cdot n}{|n+r|\,|n|} \biggr) \,\psi_n,
\end{eqnarray*}
since $\lambda(r,n)\,\bar\lambda(r,n+r) = |\lambda(r,n)|^2 = 1$ (using
the antisymmetry of~$\theta$). Similar formulas obtain for
$[F,u^r]^* [F,u^r] \psi'_n$. Thus
\begin{eqnarray*}
\|[F,u^r]\|_p^p
&=& \snorm{[F,u^r]}_p^p = 2^{1 + p/2} \sum_n
  \biggl( 1 - \frac{(n+r)\cdot n}{|n+r|\,|n|} \biggr)^{\!p/2}
\\
&=& 2 \sum_n \biggl( \frac{|r|^2}{|n|^2} - \frac{(r\cdot n)^2}{|n|^4}
                    + O(|n|^{-3})  \biggr)^{p/2},
\end{eqnarray*}
so that $[F,u^r] \in {\cal L}^p$ if and only if
$\sum_{n\neq 0} |n|^{-p}$ converges if and only if
$\int_1^\infty \rho^{2-p} \,d\rho$ converges if and only if $p > 3$.

For the general case $A = [F,X]$, the matrix of $A^*A$ has
off-diagonal terms, but one generally finds that $\snorm{[F,X]}_p^p$
diverges for $p \leq 3$, so that $[F,X] \notin {\cal L}^3$. But we can
show, with the same type of arguments, that $[F,X] \in {\cal L}^4$ for
any $X = a_r u^r \in {\Bbb T}_\theta^3$. Since
\begin{eqnarray*}
\lefteqn{[F,X]^* [F,X] \psi_n}
\\
&=& \sum_{r,s} \bar\lambda(r,s) \, \bar a_r a_s
    \biggl( \frac{(n+s)\cdot\sigma}{|n+s|}
	        - \frac{(n-r+s)\cdot\sigma}{|n-r+s|} \biggr)
\\
&& \quad\times\quad \biggl( \frac{(n+s)\cdot\sigma}{|n+s|}
	        - \frac{n\cdot\sigma}{|n|} \biggr)  \,\psi_{n-r+s},
\end{eqnarray*}
and $\|[F,X]\|_4^4 = \|B\|_2^2 = \sum_n\|B\psi_n\|^2 + \|B\psi'_n\|^2$
with $B = [F,X]^* [F,X]$, and also since
$\|(p\cdot\sigma)(q\cdot\sigma)\psi_n\|^2
 + \|(p\cdot\sigma)(q\cdot\sigma)\psi'_n\|^2 = 2|p|^2 |q|^2$, we
obtain, after replacing $n$ by $n - s$,
\begin{eqnarray*}
\lefteqn{\|[F,X]\|_4^4}
\\
&=& 2 \sum_{n,r,s} |\bar\lambda(r,s)\, \bar a_r a_s|^2
  \biggl| \frac{n}{|n|} - \frac{(n-r)}{|n-r|} \biggr|^2 \,
  \biggl| \frac{n}{|n|} - \frac{(n-s)}{|n-s|} \biggr|^2
\\
&=& 8 \sum_{n,r,s} |\bar a_r a_s|^2
  \biggl( 1 - \frac{n\cdot(n-r)}{|n|\,|n-r|} \biggr) \,
  \biggl( 1 - \frac{n\cdot(n-s)}{|n|\,|n-s|} \biggr)
\\
&=& 2 \sum_{n,r,s} \biggl( \frac{|r|^2}{|n|^2}
            - \frac{(r\cdot n)^2}{|n|^4}\biggr) |a_r|^2
  \, \biggl( \frac{|s|^2}{|n|^2}
            - \frac{(s\cdot n)^2}{|n|^4}\biggr) |a_s|^2
\\
&&\qquad\qquad{} + O(|n|^{-5}),
\end{eqnarray*}
which converges since $|r|\,a_r$ is a square-summable sequence because
$a \in {\cal S}({\Bbb Z}^3)$. Thus the quantum dimension of
${\Bbb T}_\theta^3$ is~$3$.

Results of this kind are independent of the torus
parameters~$\theta_{jk}$, so from the dimensional standpoint the
ultraviolet behaviour is exactly the same for all $3$-tori,
commutative or not.

\section{The noncommutative Chern character theorem}

We have seen that for NC tori, the quantum dimension, as measured by
the character given by the phase operator~$F$, equals the quantum
dimension. (It should be clear that the calculations for $n = 3$ yield
analogous results for higher odd~$n$.) What is the underlying reason
for this?

One of the deepest results in noncommutative geometry is that the
noncommutative integral defined by a generalized Dirac operator $D$
and the character given by its phase operator $F$ have the same values
on ``volume forms''. This is the content of Connes' Hauptsatz
\cite[p.~308]{Book}: if $n$ is odd,
\begin{eqnarray}
\lefteqn{\mathop{\rm Tr}\nolimits_C
   \Bigl( \mathop{\textstyle\sum}\nolimits_j
    a_0^j\,[F,a_1^j]\dots [F,a_n^j] \Bigr)}
\nonumber \\
&&\quad = \ncint \mathop{\textstyle\sum}\nolimits_j
       a_0^j\,[D,a_1^j]\dots [D,a_n^j] \,ds^n,
\label{eq:Hauptsatz}
\end{eqnarray}
whenever $\sum_j a_0^j \otimes a_1^j \otimes\cdots\otimes a_n^j$ is a
Hochschild $n$-cycle on the algebra~$\cal A$.

Assume that the classical dimension of a spin geometry is~$n$, and
that Hochschild cohomology of~$\cal A$ is the dual of its Hochschild
homology. If the cohomological dimension of the character (what we
have called the ``quantum dimension'' of the geometry) were lower, say
$(n - 2)$ ---it must still be an odd integer--- then the character
$\tau_n$ would necessarily \cite[p.~294]{Book} be of the
form $(-2/n)\, S\tau_{n-2}$, where $\tau_{n-2}$ is the analogous
character in degree $(n - 2)$ and the periodicity operator $S$
promotes cyclic $(n-2)$-cocycles to cyclic $n$-cocycles. However,
promoted cyclic cocycles are always Hochschild-cohomologous to zero;
if $\bbox{c}$ denotes the cycle whose representative on~$\cal H$
fulfils $\pi_D(\bbox{c}) = 1$ (Axiom~3 in Section~2), we would have
then $\ncint \,ds^n = \ncint \pi_D(\bbox{c}) \,ds^n
 = (-2/n)\, S\tau_{n-2}(\bbox{c}) = 0$, which is not possible in
classical dimension~$n$. In fine, the quantum dimension is not lower
than~$n$.

By direct computation, Langmann found~\cite{LangmannNCI}, for the usual
spin geometry on~${\Bbb R}^n$, that the character determined by the
phase operator $F = D\mkern-11.5mu/\,/|D\mkern-11.5mu/\,|$ is given,
up to a constant factor, by an ordinary de~Rham integral:
\[
\mathop{\rm Tr}\nolimits_C (a_0\,[F,a_1]\dots [F,a_n])
 = \widetilde C_n\int_{{\Bbb R}^n} \mathop{\rm tr}(a_0\,da_1\dots da_n)
\]
of smooth, compactly supported matrix-valued functions
on~${\Bbb R}^n$. (The constant $\widetilde C_n$ and the $C_n$ of
Section~2 differ only by a factor of modulus~$1$.) The integral on the
right hand side is in fact a {\it noncommutative\/} integral, due to
the trace theorem of Connes~\cite{ConnesAct}: on a spin manifold~$M$,
the following identity holds:
\begin{equation}
\ncint a_0\,[D\mkern-11.5mu/\,,a_1]\dots[D\mkern-11.5mu/\,,a_n]\,ds^n
 = \widetilde C_n \int_M a_0 \,da_1 \dots \,da_n.
\label{eq:commint}
\end{equation}
This is proved for compact manifolds by use of the Wodzicki
residue~\cite{Sirius,Kassel,ConnesAct}. Our results in~\cite{Odysseus}
extend the validity of~(\ref{eq:commint})
to~${\Bbb R}^n$, for compactly supported functions. Therefore, in the
commutative case, the integral identity~(\ref{eq:Hauptsatz}) subsumes
the formula given by Langmann.

To summarize, the Fredholm character and the integral give equal
results when evaluated on a volume form. Commutativity has nothing
to do with the matter ---except to allow the noncommutative integral
to be rewritten as an ordinary integral.

While a proof of~(\ref{eq:Hauptsatz}) is not given in~\cite{Book}, it
is a special case of an even more general index theorem proved
in~\cite{ConnesMIndex}. Thus, at the very heart of NCG, there is a
barrier to the improvement of ultraviolet behaviour by abandoning
locality of the fields. This is perhaps not a bad thing, given that
spacetime behaves at long distances as a commutative manifold of fixed
dimension. Of course, time is still counted as a $c$-number here, both
before and after quantization. It may still happen that in fully
interacting theories, the noncommutativity of space introduces
couplings that soften the ultraviolet divergences. At any rate, we
expect fermion fields over noncommutative spaces ---in particular over
Kronecker foliation algebras, that may prove the more pertinent ones
in M-theory--- to be regularizable by a direct generalization of the
methods developed in~\cite{MLSaga}, which go beyond the $1+1$ case; to
our mind, this is one of the outstanding issues.

\section{Conclusion}

Quantization, in the Hamiltonian formalism, amounts to substituting
$q$-numbers for the canonical varia\-bles. Connes' mathematical theory
leads to consider \mbox{$nc$-numbers} generalizing $c$-numbers,
probing singular geo\-me\-tries (in fact, one can argue that the
Standard Model encodes the true, noncommutative geometry of the
world~\cite{ConnesLott,IochumS,Cordelia}). We have shown a
conceptually consistent way of making $nc$-numbers into $q$-numbers.
This points to a fusion of quantum field theory and geometry, and
promises to widen the present-day scope of both.

\acknowledgments

The basic ideas of this paper came out of a exchange of views with
C.~P.~Mart\'{\i}n; we are greatly indebted to him. We are also
indebted to H.~Figueroa for a careful reading of the manuscript,
to B.~Iochum for indicating an obscurity in an earlier version, and to
E.~Alvarez and C.~G\'omez for illuminating discussions.


\begin{references}

\bibitem{ConnesDS}
A. Connes, M. R. Douglas and A. Schwartz,
hep-th/9711162.

\bibitem{LizziS}
F. Lizzi and R. J. Szabo,
hep-th/9712206.

\bibitem{ConnesAxioms}
A. Connes,
lectures given at the course of the Coll\`ege de France, Winter--Spring
1996;
A. Connes,
Commun. Math. Phys. {\bf 182}, 155 (1996);
A. Connes,
J. Geom. Phys. {\bf 23}, 206 (1997).

\bibitem{ConnesIHES}
A. Connes,
Publ. Math. IHES {\bf 62}, 257 (1985).

\bibitem{ConnesGNC}
A. Connes,
{\it G\'eom\'etrie Non Commutative\/} (Inter\'editions, Paris, 1990).

\bibitem{Book}
A. Connes,
{\it Noncommutative Geometry\/} (Academic Press, London, 1994).

\bibitem{ArakiB}
H. Araki,
Contemp. Math. {\bf 62}, 23 (1987).

\bibitem{ArakiS}
H. Araki,
``Schwinger terms and cyclic cohomology'',
in {\it Quantum Theories and Geo\-me\-try}, M. Cahen and M. Flato,
eds. (Kluwer, Dordrecht, 1988), p.~1.

\bibitem{MLSaga}
E. Langmann and J. Mickelsson,
Phys. Lett. B {\bf 338}, 241 (1994);
E. Langmann,
Commun. Math. Phys. {\bf 162}, 1 (1994);
J. Mickelsson,
``Wodzicki residue and anomalies of current algebras'',
hep-th/9404093;
J. Mickelsson,
``Schwinger terms, gerbes and operator residues'', hep-th/9509002;
E. Langmann and J. Mickelsson,
J. Math. Phys. {\bf 37}, 3933 (1996);
E. Langmann,
Acta Phys. Polon. B {\bf 27}, 2477 (1996);
E. Langmann,
J. Geom. Phys. {\bf 22}, 259 (1997).

\bibitem{Kreimer}
D. Kreimer,
``On the Hopf algebra structure of perturbative quantum field
theories'', q-alg/9707029.

\bibitem{ConnesMHopf}
A. Connes, talk at the II Workshop on Noncommutative Geometry and
Fundamental Interactions, Vietri-sul-Mare, March 1998;
A. Connes and H. Moscovici,
``Hopf algebras, cyclic cohomology and the transverse fundamental
class'', forthcoming.

\bibitem{BaezSZ}
J. C. Baez, I. E. Segal and Z. Zhou,
{\it Introduction to Algebraic and Constructive Quantum Field Theory\/}
(Princeton Univ. Press, Princeton, NJ, 1992).

\bibitem{Rhea}
J. M. Gracia-Bond\'{\i}a and J. C. V\'arilly,
J. Math. Phys. {\bf 35}, 3340 (1994).

\bibitem{PressleyS}
A. Pressley and G. Segal,
{\it Loop Groups\/} (Clarendon Press, Oxford, 1986).

\bibitem{Fuchs}
J. Fuchs,
``Lectures on conformal field theory and Kac-Moody algebras'',
hep-th\slash 9702194.

\bibitem{HaagKastler}
R. Haag and D. Kastler,
J. Math. Phys. {\bf 5}, 848 (1964).

\bibitem{Jordan}
P. Jordan,
Z. Phys. {\bf 44}, 473 (1927);
P. Jordan and O. Klein,
Z. Phys. {\bf 45}, 751 (1927);
P. Jordan and E. P. Wigner,
Z. Phys. {\bf 47}, 631 (1928).

\bibitem{Marshak}
R. E. Marshak,
{\it Conceptual Foundations of Modern Particle Physics},
World Scientific, Singapore, 1993.

\bibitem{Portia}
J. C. V\'arilly,
``An introduction to noncommutative geometry'', lectures at EMS
Summer School on Noncommutative Geometry and its Applications,
Monsaraz and Lisboa, 1997; physics/9709045.

\bibitem{KadisonR}
R. V. Kadison and J. R. Ringrose,
{\it Fundamentals of the Theory of Operator Algebras II\/}
(Academic Press, Orlando, FL, 1986).

\bibitem{Bost}
J.-B. Bost,
Invent. Math. {\bf 101}, 261 (1990).

\bibitem{Sirius}
J. C. V\'arilly and J. M. Gracia-Bond{\'\i}a,
J. Geom. Phys. {\bf 12}, 223 (1993).

\bibitem{Kassel}
C. Kassel,
``Le r\'esidu noncommutatif'',
S\'eminaire Bourbaki {\bf 708} (1989).

\bibitem{CiprianiGS}
F. Cipriani, D. Guido and S. Scarlatti,
J. Oper. Theory {\bf 35}, 179 (1996).

\bibitem{Plymen}
R. J. Plymen,
J. Oper. Theory {\bf 16}, 305 (1986).

\bibitem{FreedU}
D. S. Freed and K. K. Uhlenbeck,
{\it Instantons and Four-Manifolds\/}
(Springer, Berlin, 1984).

\bibitem{RieffelP}
M. A. Rieffel,
Can. J. Math. {\bf 40}, 257 (1988).

\bibitem{RieffelS}
M. A. Rieffel and A. Schwarz,
``Morita equivalence of multidimensional noncommutative tori'',
math/9803057.

\bibitem{Emch}
G. G. Emch,
{\it Algebraic Methods in Statistical Mechanics and Quantum Field
Theory\/} (Wiley, New York, 1972).

\bibitem{ShaleS}
D. Shale and W.~F. Stinespring,
J. Math. Mech. {\bf 14}, 315 (1965).

\bibitem{Odysseus}
R. Estrada, J. M. Gracia-Bond\'{\i}a and J. C. V\'arilly,
Commun. Math. Phys. {\bf 191}, 219 (1998).

\bibitem{LangmannS}
E. Langmann and G. W. Semenoff,
Phys. Lett. B {\bf 341}, 195 (1994).

\bibitem{MickelssonR}
J. Mickelsson and S. G. Rajeev,
Commun. Math. Phys. {\bf 116}, 365 (1988).

\bibitem{GohbergM}
I. C. Gohberg and A. S. Markus,
Mat. Sbornik {\bf 64} 481 (1964): AMS Translations {\bf 52}, 201 (1966).

\bibitem{LangmannNCI}
E. Langmann,
J. Math. Phys. {\bf 36}, 3822 (1995).

\bibitem{ConnesAct}
A. Connes,
Commun. Math. Phys. {\bf 117}, 673 (1988).

\bibitem{ConnesMIndex}
A. Connes and H. Moscovici,
Geom. Func. Anal. {\bf 5}, 174 (1995).

\bibitem{ConnesLott}
A. Connes and J. Lott,
Nucl. Phys. B (Proc. Suppl.) {\bf 18}, 29 (1990).

\bibitem{IochumS}
B. Iochum and T. Sch\"ucker,
Commun. Math. Phys. {\bf 178}, 1 (1996).

\bibitem{Cordelia}
C. P. Mart\'{\i}n, J. M. Gracia-Bond\'{\i}a and J. C. V\'arilly,
Phys. Reports {\bf 294}, 363 (1998).

\end{references}
\end{document}